\documentclass{article}
\usepackage{arxiv}

\usepackage{amsmath,amsfonts,array,textcomp,stfloats,url,verbatim,graphicx, cite,multirow,hyperref,algorithm,booktabs,xcolor}
\usepackage[caption=false,font=normalsize,labelfont=sf,textfont=sf]{subfig}
\usepackage[noend]{algorithmic}

\begin{document}
\title{LLM-Assisted Light: Leveraging Large Language Model Capabilities for Human-Mimetic Traffic Signal Control in Complex Urban Environments}

\author{Maonan Wang \\
	The Chinese University of Hong Kong, Shenzhen, China \\
    Shanghai AI Laboratory, Shanghai, China \\
	\texttt{maonanwang@link.cuhk.edu.cn} \\
	\And
	Aoyu Pang \\
	The Chinese University of Hong Kong, Shenzhen, China \\
	\texttt{aoyupang@link.cuhk.edu.cn} \\
	\And
	Yuheng Kan \\
	SenseTime Group Limited, Shanghai, China \\
	Shanghai AI Laboratory, Shanghai, China \\
	\texttt{kanyuheng@sensetime.com} \\
	\And
	Man-On Pun \\
	The Chinese University of Hong Kong, Shenzhen, China \\
	\texttt{simonpun@cuhk.edu.cn} \\
	\And
	Chung Shue Chen \\
	Nokia Bell Labs, Paris, France \\
	\texttt{chung\_shue.chen@nokia-bell-labs.com} \\
    \And
	Bo Huang \\
	The University of Hong Kong, Hongkong, China \\
	\texttt{bohuang@hku.hk} \\
}

\maketitle
\renewcommand{\shorttitle}{LLM-Assisted Light (LA-Light)}

\begin{abstract}
Traffic congestion in metropolitan areas presents a formidable challenge with far-reaching economic, environmental, and societal ramifications. Therefore, effective congestion management is imperative, with traffic signal control (TSC) systems being pivotal in this endeavor. Conventional TSC systems, designed upon rule-based algorithms or reinforcement learning (RL), frequently exhibit deficiencies in managing the complexities and variabilities of urban traffic flows, constrained by their limited capacity for adaptation to unfamiliar scenarios. In response to these limitations, this work introduces an innovative approach that integrates Large Language Models (LLMs) into TSC, harnessing their advanced reasoning and decision-making faculties. Specifically, a hybrid framework that augments LLMs with a suite of perception and decision-making tools is proposed, facilitating the interrogation of both the static and dynamic traffic information. This design places the LLM at the center of the decision-making process, combining external traffic data with established TSC methods. Moreover, a simulation platform is developed to corroborate the efficacy of the proposed framework. The findings from our simulations attest to the system's adeptness in adjusting to a multiplicity of traffic environments without the need for additional training. Notably, in cases of Sensor Outage (SO), our approach surpasses conventional RL-based systems by reducing the average waiting time by $20.4\%$. This research signifies a notable advance in TSC strategies and paves the way for the integration of LLMs into real-world, dynamic scenarios, highlighting their potential to revolutionize traffic management. The related code is available at \href{https://github.com/Traffic-Alpha/LLM-Assisted-Light}{https://github.com/Traffic-Alpha/LLM-Assisted-Light}.
\end{abstract}


\section{Introduction}
Traffic congestion poses a significant challenge globally, leading to adverse economic, environmental, and social impacts \cite{sweet2011does}. Managing traffic flow efficiently, especially at road intersections, is crucial to alleviate congestion. Traffic signal control (TSC) systems are vital in this effort \cite{zhao2011computational}. Traditional rule-based TSC methods, such as the Webster method \cite{koonce2008traffic} and Self-Organizing Traffic Light Control (SOTL) \cite{cools2013self}, have been somewhat effective in managing traffic flow and reducing congestion. Yet, these systems are inherently limited by their static, rule-based algorithms that do not fully adapt to the ever-changing patterns of urban traffic \cite{qadri2020state}.

Recently, the evolution of sensor technology and data collection has led to the development of more adaptive TSC strategies. In particular, Reinforcement Learning (RL) has emerged as an attractive approach, utilizing real-time data to dynamically adjust traffic signals \cite{wei2019survey}. Despite their potentials, these RL-based TSC systems are not without limitations. These systems may suffer from overfitting to specific traffic patterns. Additionally, RL systems typically rely on reward functions that may not be able to capture infrequent but critical events, such as emergency vehicles' sudden arrivals or unexpected road blockages. This can reduce their practicality in real-world conditions \cite{vardhan2021rare}.

In response to these limitations, this paper introduces a novel approach that integrates Large Language Models (LLMs) into the TSC framework to assist in the decision-making process, named \textbf{L}LM-\textbf{A}ssist Light (LA-Light). Our method leverages the extensive knowledge and ``common sense'' reasoning abilities of LLMs to enhance decision-making in complex and uncommon traffic situations. LLMs, with their sophisticated natural language processing capabilities, can interpret intricate traffic scenarios and recommend actions that may be overlooked by rule-based or RL-based systems. Furthermore, we introduce a set of tools specifically designed to bridge the gap between the TSC system and the LLM. These tools act as intermediaries, collecting environmental data and communicating with the LLM, which then guides the TSC system. Additionally, they feature a standardized interface, making LA-Light compatible with existing TSC methods, and enabling it to serve as an auxiliary tool within the LA-Light framework. This collaborative process allows for a well-rounded control strategy that not only makes informed decisions but also provides justifications for these decisions, thus improving the transparency of the system and building trust with traffic management operators.

\begin{figure}[!t]
    \centering
    \includegraphics[width=0.5\linewidth]{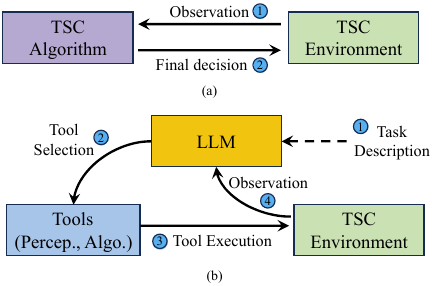}
    \caption{Comparative framework analysis between LA-Light and conventional TSC systems. (a) illustrates a conventional TSC system wherein decisions are made directly by an algorithm that processes environmental inputs. (b) depicts the proposed LA-Light framework, which employs an LLM for the task of traffic signal control. In LA-Light, the LLM begins by selecting the most relevant tools from an enhanced set, including perception tools and decision-making algorithms, to collect and analyze traffic data. It then evaluates the information, adjusting its choice of tools as needed, until a definitive traffic control decision is formulated.}
    \label{fig:compare}
\end{figure}

Fig.~\ref{fig:compare} shows the difference between the method proposed in this paper and existing signal light control methods. The existing TSC systems, as shown in Fig.~\ref{fig:compare}a, operate by making decisions based on predefined rules and observations, which may not suffice in unusual or unpredictable events. In contrast, our approach, depicted in Fig.~\ref{fig:compare}b, integrates an LLM Agent Module to enhance the system's comprehension of various traffic scenarios and the logic behind its decisions. Given that LLMs intrinsically lack the capacity for direct engagement with traffic ecosystems or their data, an array of enhanced tools are devised to collect both the static and dynamic traffic information, subsequently facilitating the decision-making procedure. Crucially, this ensemble integrates existing rule-based and RL-based algorithms, guaranteeing that our methodology sustains state-of-the-art (SOTA) performance under conventional traffic conditions while also adapting effectively to exceptional situations.

To corroborate the efficacy of the proposed framework, a simulation platform is developed. Extensive experiments are conducted on the simulation platform considering various intersection configurations. It is shown that the proposed LA-Light system achieves good performance in typical scenarios as well as in situations involving rare events. We also provide several qualitative examples where LA-Light accurately analyzes intricate traffic conditions and makes more reasonable decisions than conventional TSC methods. The experiments highlight the LLM-assisted system’s capability to deeply understand traffic scenarios and to provide clear explanations for its actions. LA-Light is shown to make informed decisions that enhance safety, efficiency, and comfort, outperforming existing methods that may otherwise fail or yield suboptimal results under challenging conditions. The key contributions of this study are summarized as follows.
\begin{itemize}
    \item We propose LA-Light, a hybrid TSC framework that integrates the human-mimetic reasoning capabilities of LLMs, enabling the signal control algorithm to interpret and respond to complex traffic scenarios with the nuanced judgment typical of human cognition. This innovation allows for seamless adaptation to urban traffic challenges, particularly in addressing the unpredictable and rare events that conventional systems may overlook;
    \item A closed-loop traffic signal control system has been developed, integrating LLMs with a comprehensive suite of interoperable tools. This integration yields in-depth insights into consistent and variable traffic patterns, thereby equipping the system with the capability for real-time analysis and decision-making that mirrors human intelligence. Additionally, the system is designed with a standardized interface, which allows for rapid integration with existing TSC methods and enables easy customization by users.
    \item Through comprehensive experimentation, the results show that our model is adept at understanding and responding to a variety of environmental changes. In particular, it is able to address rare or unexpected events and provide superior performance, thereby validating its practical applicability and efficacy in real-world settings.
\end{itemize}

The remainder of this paper is organized as follows: Section~\ref{sec_related_work} provides an overview of existing literature in the fields of TSC and LLMs whereas Section~\ref{sec_preliminary} defines key terminologies pertaining to TSC that are referenced throughout this paper. After that, Section~\ref{sec_method} details the architecture of the proposed LLM-Assist Light framework, encompassing the utilized tools and the construction of prompts. Section~\ref{sec_experiment} describes the experiments conducted to validate our approach. Finally, Section~\ref{sec_conclude} provides some concluding remarks and suggestions for future research directions.

\section{Related Work} \label{sec_related_work}

\subsection{Traffic Signal Control Methods}

The pursuit of effective traffic signal control (TSC) strategies in urban settings is a well-established challenge with the goal of alleviating congestion. The rule-based TSC methods have been designed to optimize traffic signals under a variety of traffic conditions \cite{martinez2011survey}. For example, the Webster method \cite{koonce2008traffic} calculates the ideal cycle length and distribution of traffic signal phases at intersections, based on traffic volumes and the assumption of a stable flow of traffic over a specific period. The Self-Organizing Traffic Light Control (SOTL) scheme \cite{cools2013self} uses a set of predetermined rules to decide whether to continue with the current traffic signal phase or to change it. Adaptive TSC systems such as the Split Cycle Offset Optimization Technique (SCOOT) \cite{hunt1982scoot} and the Sydney Coordinated Adaptive Traffic System (SCATS) \cite{lowrie1990scats} dynamically alter cycle lengths, phase divisions, and offsets by choosing from a collection of predefined plans in response to live traffic sensor data. Although conventional TSC methods have achieved some success in mitigating congestion, their effectiveness is hindered by limitations in real-time traffic data utilization and difficulties in adapting to quickly changing traffic situations. Moreover, these methods often fall short in complex traffic scenarios \cite{wei2019survey}.

Recent advancements in TSC have seen a shift towards RL-based systems, which are increasingly favored for their ability to dynamically manage traffic lights \cite{wu2023deep}. These systems typically use factors such as queue length \cite{wei2019presslight,zang2020metalight,chen2020toward,pang2024scalable,gu2024large}, vehicle waiting time \cite{chu2019multi,wang2022adlight, wang2023unitsa} or intersection pressure \cite{varaiya2013max, wei2019presslight, oroojlooy2020attendlight,jiang2024general} as key indicators in their reward functions, training agents to reduce congestion. Furthermore, the frequency of signal switching has been considered \cite{wei2018intellilight} to prevent the negative impacts of rapid signal changes, such as increased stop-and-go driving and the risk of accidents. Although RL-based TSC systems offer flexibility in optimizing traffic flow by adjusting the reward function, finding the right balance for these factors is a complex task \cite{bouktif2023deep}. Furthermore, if the reward function does not encompass infrequent but critical events, it may not provide the agent with sufficient direction to handle unexpected conditions effectively \cite{vardhan2021rare}.

\subsection{Large Language Models}

Large language models (LLMs), such as the Generative Pretrained Transformer (GPT) series \cite{floridi2020gpt}, including its advanced versions like GPT-3.5 and GPT-4 Turbo \cite{chatgpt2023}, as well as open-source counterparts like Llama \cite{touvron2023llama} and Llama2 \cite{touvron2023llama2}, constitute a category of artificial intelligence systems designed to understand, generate and modify human language. These models rely on complex machine learning algorithms, specifically the transformer architecture \cite{vaswani2017attention}, and are trained on extensive text datasets. Such comprehensive training grants them a sophisticated grasp of language nuances. A specialized version, InstructGPT \cite{ouyang2022training}, has been further refined to interpret user instructions with greater precision, providing relevant responses in a range of applications, from creating content to retrieving information. The "Chain-of-Thought" reasoning method \cite{wei2022chain} has introduced improved functionality within LLMs, enhancing their capability to solve complex problems by processing a series of logical steps. This significantly improves their performance in tasks that require arithmetic, commonsense, and symbolic reasoning. The "ReAct" strategy \cite{yao2022react} extends the abilities of LLMs in complex tasks that require both reasoning and decision making. By prompting models to alternate between verbal reasoning and action generation, this approach enables dynamic reasoning and interaction with external systems.

Motivated by their superior performance, LLMs have been recently investigated for a multitude of tasks, as documented in recent surveys \cite{zhao2023survey, xi2023rise}. These models have found applications in transportation planning and decision support systems \cite{cui2024survey}. For instance, \cite{cui2023receive} involves LLMs as decision-makers in conjunction with perception and positioning systems to aid autonomous vehicles whereas the Open-TI framework \cite{da2023open} integrates LLMs with traffic analysis tools, facilitating complex command execution through natural language interactions. Furthermore, LLMs have also been utilized to enhance lane-changing maneuvers in vehicles, aligning them more closely with human-like decision-making \cite{fu2024drive}. Finally, there are a limited number of studies have explored the application of LLMs in controlling traffic signals \cite{lai2023large, tang2024large}. These studies primarily focus on standard scenarios and lack compatibility with existing TSC systems. In contrast, the method proposed in this paper extends the use of LLMs to manage traffic signals across both normal and abnormal traffic conditions. Moreover, by positioning the LLM as the central decision-maker that invokes various tools, the LA-Light framework achieves seamless compatibility with established TSC methods.

\section{Preliminaries} \label{sec_preliminary}

\begin{figure}[!ht]
    \centering
    \includegraphics[width=0.5\linewidth]{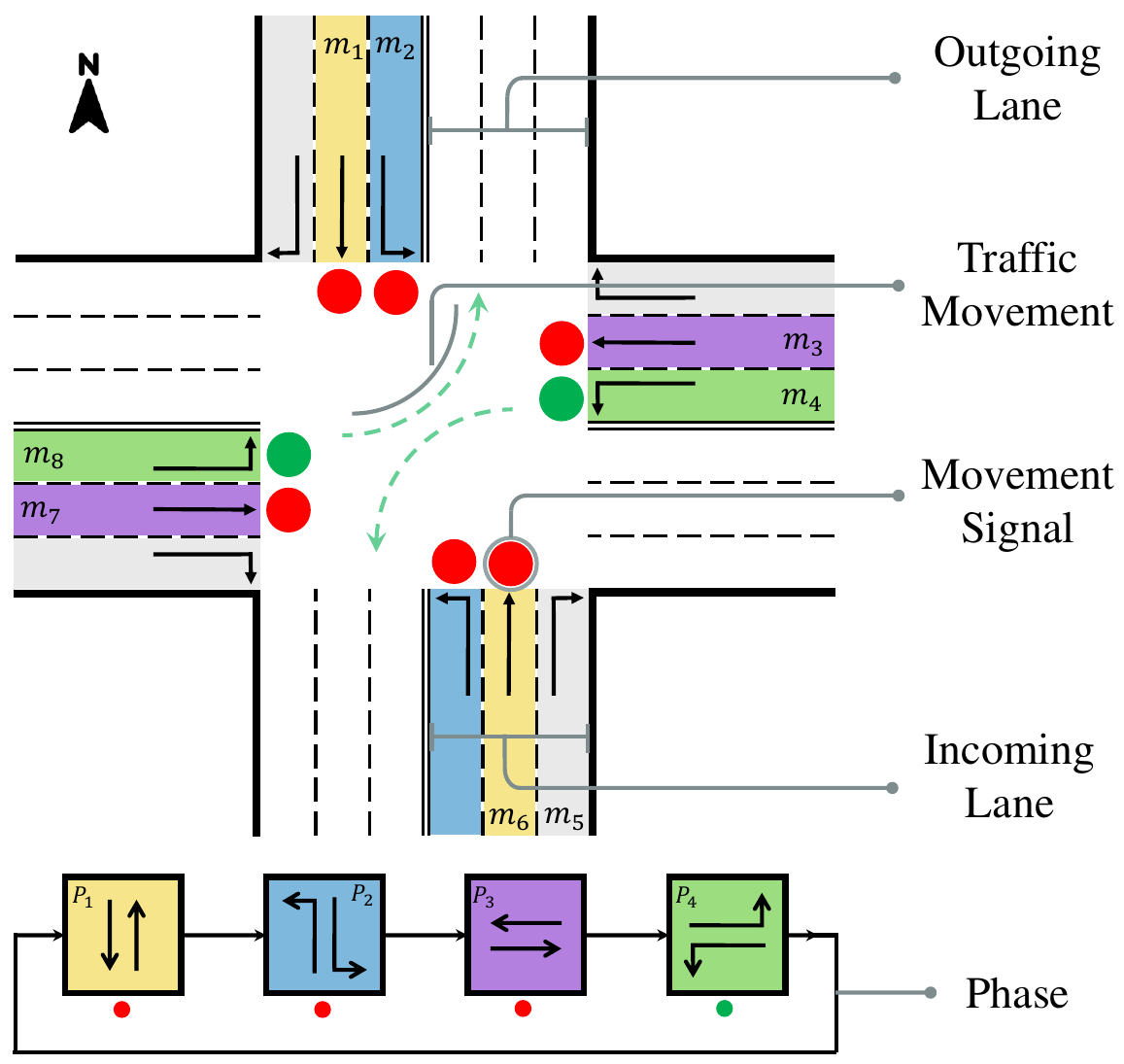}
    \caption{A standard four-legged road intersection with eight traffic movements and signal phases.}
    \label{fig:preliminaries}
\end{figure}

This section provides definitions for the key terminologies about TSC that are utilized throughout this paper. These terms are described with reference to a typical four-legged intersection, which is illustrated in Fig.~\ref{fig:preliminaries}.

\textbf{Lanes}: 
In the context of traffic intersections, lanes are categorized according to their purpose for the intersection. \textit{Incoming lanes} guide vehicles toward the intersection, allowing them to enter. Conversely, \textit{outgoing lanes} are designed to allow vehicles to depart from the intersection.

\textbf{Traffic Movements}: 
The directional flow of vehicles from incoming to outgoing lanes defines traffic movement at an intersection. A conventional four-way intersection typically includes four directional paths: East (E), West (W), North (N), and South (S). Each path facilitates two main vehicular movements for exiting: a left turn, indicated by $l$, and a straight-ahead movement, denoted by $s$. In the context of this study, right-turn movements are excluded on the assumption that they are not signal-controlled in regions with right-hand traffic systems. Therefore, the traffic control system at the intersection administers eight distinct movements, labeled as $m_1$ through $m_8$.

\textbf{Movement Signals}: 
Movement signals are the controls that dictate the flow of traffic for each direction at an intersection. A green signal authorizes the traffic to proceed, while a red signal indicates a stop condition. As depicted in Fig.~\ref{fig:preliminaries}, at a standard four-way intersection, if movement signals $m_4$ and $m_8$ are green, this indicates that the corresponding movements—specifically, the westbound and eastbound left turns—are allowed, and all other movements are halted.

\textbf{Phases}: 
A traffic signal phase is a combination of movement signals that are displayed at the same time to manage multiple traffic flows. This configuration is designed so that all movements within a phase can proceed safely and without interference from other directions. As shown in Fig.~\ref{fig:preliminaries}, the intersection operates using four distinct phases, identified as $P_1$ through $P_4$. For example, during phase $P_4$, the green signals for movements $m_4$ and $m_8$ are activated concurrently, which enables vehicles to make left turns from both the westbound and eastbound approaches without conflict.

\section{LLM-Assisted Light} \label{sec_method}

\begin{figure*}[!ht]
    \centering
    \includegraphics[width=0.99\linewidth]{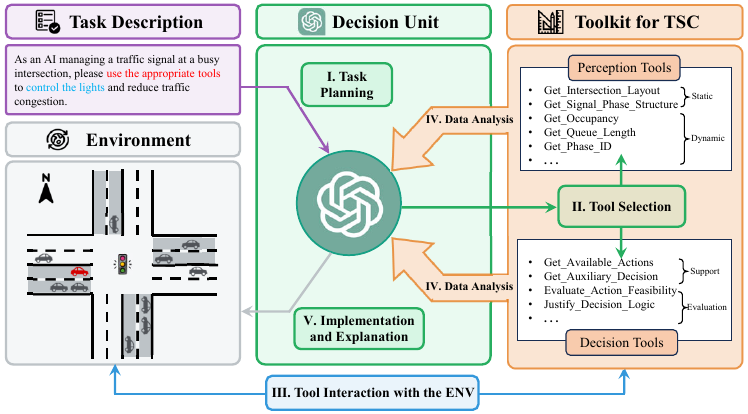}
    \caption{The LA-Light framework: A schematic representation of the five-step process for integrating LLM in TSC. \textbf{Step 1} outlines the task planning phase where the LLM defines its role in traffic management. \textbf{Step 2} involves the selection of appropriate perception and decision-making tools by the LLM. In \textbf{Step 3}, these tools interact with the traffic environment to gather data. \textbf{Step 4} depicts the analysis of this data by the Decision Unit to inform decision-making. Finally, \textbf{Step 5} illustrates the implementation of the LLM's decisions and the provision of explanatory feedback for system transparency and validation.}
    \label{fig:framework}
\end{figure*}

\subsection{Overview of LA-Light}

The LA-Light framework introduces an innovative hybrid decision-making process for TSC that leverages the cognitive capabilities of LLMs alongside traditional traffic management methodologies. As illustrated in Fig.~\ref{fig:framework}, the framework operates through a sequence of five methodical steps for decision-making, commencing with the specification of the LLM's role. In this initial phase, the LLM is assigned the function of regulating traffic signals at busy intersections to alleviate congestion, drawing on a combination of analytical and control tools.

Following this, the LLM is responsible for choosing the most appropriate tools from a predefined set of tools. These tools are divided into two categories, namely the perception tools and the decision-making tools. The perception tools are tasked with collecting a range of environmental data, both dynamic and static, to form a detailed picture of the traffic conditions. Conversely, the decision-making tools are specifically engineered for facilitating decisions and can be further classified into two categories: decision support tools, which utilize extant TSC algorithms to aid the decision-making process, and decision verification tools that assess the precision of decisions rendered by the LLM.

In the third stage, the chosen tools are activated within the traffic environment to collate traffic data, which is critical for informed decision-making. The collected data are then conveyed to the LLM, including the chat history, constituting the fourth step. At this point, the LLM scrutinizes the data to determine the next course of action. It evaluates the adequacy of the current data set and determines whether there is a necessity to activate supplementary tools for enhanced data acquisition.

Once sufficient data are obtained, the LLM proceeds to formulate traffic signal timing recommendations. These recommendations are then transmitted to the traffic control systems, implementing the recommended adjustments to the traffic light phases. The specific action taken in this study is the selection of an appropriate traffic phase ID for the junction, which the traffic lights then adopt. Concurrently, the LLM elucidates the reasoning behind its recommendations, thus improving the system's transparency and intelligibility. This aspect is vital for traffic operators, as it bolsters the reliability and trust in the system's operations. 

The discussions above delineate the decision-making process at each juncture. In Algorithm~\ref{algo:la-light}, we introduce an elaborate control sequence that incorporates several decision-making cycles. Each cycle embodies the five steps previously outlined. Furthermore, the content of the dialogue is preserved in the context dialogue memory ($M$), which allows the LLM to integrate contextual data and construct a logical sequence for future decisions.

\begin{algorithm}[!ht]
\caption{LA-Light: LLM-Assisted Traffic Signal Control}
\label{algo:la-light}
\begin{algorithmic}[1]
\small
\STATE \textbf{Input}: Total simulation time $T$, current time $t$, Intervention Frequency $\Delta t$, task description $D_{\text{task}}$, tool $x \in \mathcal{X}$, context dialogue memory $M$
\STATE \textbf{Initialize}: $t = 0$, $M = []$
\WHILE{$t < T$}
    \STATE \texttt{// Append task description to memory}
    \STATE $M.\text{append}(D_{\text{task}})$
    \STATE $done = \text{False}$
    \WHILE{not $done$}
        \STATE \texttt{// Choose the appropriate tool}
        \STATE $x = \texttt{LLM}(M)$

        \STATE \texttt{// Interact with environment}
        \STATE $obs = x(\text{env})$

        \STATE \texttt{// Record the tool used and observation}
        \STATE $M.\text{append}((x, obs))$

        \STATE \texttt{// Determine if decision can be made}
        \STATE $done = \text{DecisionCriterion}(obs)$
    \ENDWHILE

    \STATE \texttt{// Execute decision in environment}
    \STATE $\text{env}.\text{execute}(\text{decision})$
    \STATE $t = t + \Delta t$
\ENDWHILE
\STATE \textbf{Output}: Implemented traffic signal timing adjustments and rationale
\end{algorithmic}
\end{algorithm}

\begin{figure*}[!ht]
    \centering
    \includegraphics[width=0.99\linewidth]{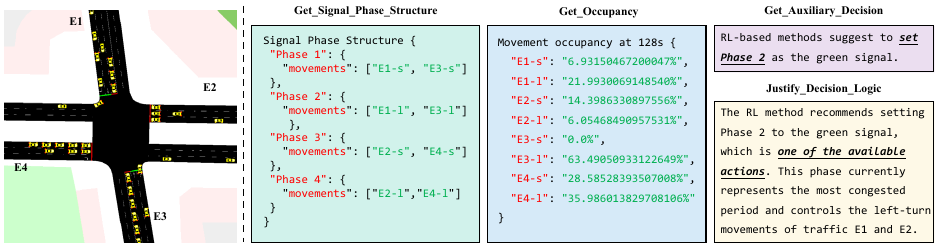}
    \caption{Examples of tools in the LA-Light framework. Two perception tools, \textit{Get\_Signal\_Phase\_Structure} and \textit{Get\_Occupancy}, allow users to obtain the traffic signal phase and congestion level of the intersection, respectively. Two decision tools, \textit{Get\_Auxiliary\_Decision} provide the decision of the RL-based method, and \textit{Justify\_Decision\_Logic} explains the decision according to the current situation of the intersection.}
    \label{fig:llm_tools_example}
\end{figure*}

\subsection{Toolkit for TSC}

The LA-Light framework incorporates a comprehensive set of tools that facilitate the interaction of LLMs with the traffic environment. These tools, acting as sensory and cognitive extensions, enable the LLMs to accurately perceive traffic conditions and make well-informed decisions. The toolkit is divided into two main categories: \textbf{Perception Tools} and \textbf{Decision Tools}. Perception Tools are focused on the acquisition of static and dynamic traffic information, while Decision Tools support and evaluate the decision-making process. The design of these tools is modular and scalable, ensuring easy integration of new functionalities to accommodate various traffic management challenges. This approach allows the LLM to effectively combine traditional traffic control methods with its advanced reasoning capabilities, improving its performance in complex traffic scenarios. 

Fig.~\ref{fig:llm_tools_example} illustrates the effects of some tools. For a crossroad, perception tools can obtain static signal light phase information or dynamic traffic movement shares. In Fig.~\ref{fig:llm_tools_example}, we obtained the current intersection's four traffic phases and the traffic movements contained in each phase. At the same time, the congestion level of each traffic movement at this moment is obtained through the tool \textit{Get\_Occupancy}. Similarly, decision tools are used to provide RL-based decisions, such as setting Phase-2 to green. Furthermore, \textit{Justify\_Decision\_Logic} is used to give explanations based on the current junction conditions. The following list details all the tools used in the LA-Light framework.

\textbf{Perception Tools (Static)}: The static subset of Perception Tools is responsible for capturing the unchanging aspects of the traffic environment. These include:
\begin{itemize}
    \item \textit{Get\_Intersection\_Layout}: This tool delineates the intersection's configuration, detailing the number and function of lanes associated with each direction, which is fundamental for understanding potential traffic flow scenarios.
    \item \textit{Get\_Signal\_Phase\_Structure}: This tool offers a detailed description of the traffic signal phases at the intersection, outlining the sequence and associated traffic movements for each phase.
\end{itemize}

\textbf{Perception Tools (Dynamic)}: The dynamic category of Perception Tools are responsible for gathering real-time, fluctuating traffic parameters:
\begin{itemize}
    \item \textit{Get\_Occupancy}: This function calculates the proportion of space currently occupied by vehicles in each movement, providing insights into congestion levels and the distribution of vehicles at the intersection.
    \item \textit{Get\_Queue\_Length}: It measures the length of vehicle queues for each traffic movement, providing quantitative data to gauge traffic backlogs.
    \item \textit{Get\_Phase\_ID}: This tool identifies the currently active traffic signal phase at the intersection, which is crucial for the LLM to understand which traffic flows are being allowed at any given moment.
    \item \textit{Get\_Junction\_Situation}: This tool is designed to detect and assess unusual or emergency situations at the intersection, such as the arrival of emergency vehicles or traffic accidents, that may require immediate action or a departure from standard traffic control measures.
\end{itemize}

\textbf{Decision Tools (Support)}: These tools are designed to aid the LLM in the decision-making process by offering reference points and additional insights:
\begin{itemize}
    \item \textit{Get\_Auxiliary\_Decision}: This function offers alternative decisions that can act as a reference or provide additional viewpoints to the LLM's decision-making process. Within LA-Light, the UniTSA method \cite{wang2023unitsa} is implemented as the foundational approach. UniTSA is an RL strategy that uses queue length as the reward metric. It is a universal RL-based method suitable for junctions with diverse configurations, eliminating the need for separate training for different junctions and still achieving good performance in standard situations.
    \item \textit{Get\_Available\_Actions}: This tool lists the potential actions available to the LLM at any given time, outlining the range of immediate decision-making options.
\end{itemize}

\begin{figure}[!ht]
	\centering
	\includegraphics[width=0.5\linewidth]{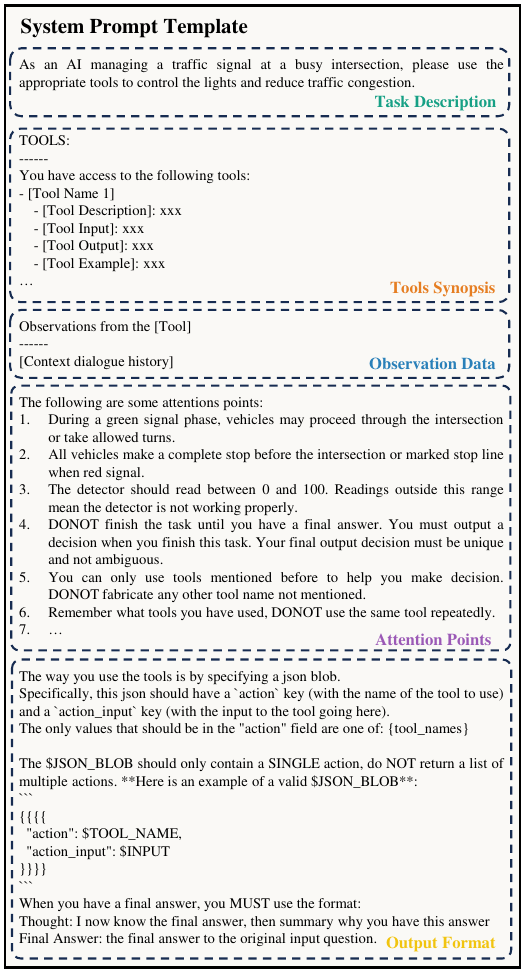}
	\caption{System Prompt structure within the LA-Light framework. The design incorporates five components: \textbf{(1) Task Description}, detailing the LLM's role in traffic signal management; \textbf{(2) Tools Synopsis}, providing a catalog and description of available traffic control tools; \textbf{(3) Observations Data}, compiling data from tool feedback and chat history of the preceding cycle; \textbf{(4) Attention Points}, emphasizing compliance with traffic regulations and safety guidelines in tool deployment; and \textbf{(5) Output Format}, defining the protocol for the LLM's decision communication to ensure proper tool utilization.}
	\label{fig:system_prompt}
\end{figure}

\textbf{Decision Tools (Evaluate)}: The evaluative subset of Decision Tools facilitates the validation and justification of the decisions proposed by the LLM:
\begin{itemize}
    \item \textit{Evaluate\_Action\_Feasibility}:This tool verifies the compatibility of the LLM-generated decisions with the expected output format, as depicted in the latter part of Fig.~\ref{fig:system_prompt}. Its primary function is to identify and mitigate the generation of irrelevant or incorrect information, known as LLM hallucinations. Should the output fail to meet the specified criteria, the LLM is prompted to revise and resubmit its decision.
    \item \textit{Justify\_Decision\_Logic}: This function allows the LLM to explain the reasoning behind its decisions. Such explanations increase the transparency and understanding of the TSC process for traffic managers. 
\end{itemize}

The toolkit described above is essential for the operation of the LA-Light framework, serving to connect the LLM's sophisticated reasoning abilities with the practical needs of TSC. Although the current set of tools is tailored to the particular requirements identified in this study, the framework's flexible architecture allows for the easy addition of new tools. This adaptability ensures that the framework can meet the changing demands of traffic management and fulfill future objectives.

\begin{table*}[!ht]
    \centering
    \caption{The details of the prompt components for \textit{Get\_Intersection\_Layout}}
    \label{tab_example_tool_prompt}
    \resizebox{\textwidth}{!}{%
    \begin{tabular}{cc}
    \hline
    \textbf{Name} & \textbf{Description} \\ \hline
    \textit{\textbf{Description}} & \begin{tabular}[c]{@{}c@{}}The Get\_Intersection\_Layout function provides a detailed configuration of an intersection's layout.\\ It returns the number and function of lanes for each direction at the specified intersection.\end{tabular} \\ \hline
    \textit{\textbf{Input}} & junction\_id (str): A string identifier for the intersection you wish to query. \\ \hline
    \textit{\textbf{Output}} & \begin{tabular}[c]{@{}c@{}}A dictionary where each key represents a traffic movement id at the intersection.\\ The corresponding value is another dictionary with the following keys:\\ - "direction": A string indicating the lane direction, where 's' is for straight, 'l' is for left turn, and 'r' is for right turn.\\ - "number\_of\_lanes": An integer representing the number of lanes for the specified direction.\end{tabular} \\ \hline
    \textit{\textbf{Example}} & \begin{tabular}[c]{@{}c@{}}To get the layout of intersection 'J1', call the function as follows:\\ layout = Get\_Intersection\_Layout('J1')\\ The expected output would be a dictionary describing the intersection layout, such as:\\ layout=\{"E1": \{"direction": "s", "number\_of\_lanes": 2\}, "E2": \{"direction": "l", "number\_of\_lanes": 1\},...\}\end{tabular} \\ \hline
    \end{tabular}}
\end{table*}

\subsection{Prompt Design}

The LA-Light system harnesses the interpretative capabilities of the LLM and refines its decision-making process through careful prompt engineering. This approach directs the model's decisions in complex traffic situations. The system prompt is meticulously crafted to encompass a broad spectrum of considerations, enabling the LLM to accurately interpret traffic conditions and effectively manage traffic signals to mitigate congestion. The system prompt comprises five components, as depicted in Fig.~\ref{fig:system_prompt}.

The initial component is the task description, which articulates the LLM's objective in traffic signal control. After the task description, the second component is a detailed briefing on the functionalities of each integrated tool within the LA-Light framework. This briefing encompasses a list of tools, their respective purposes, the types of input they require, the output they generate, and illustrative examples of their use in practice. The third component consolidates observational data, which includes the most recent outputs from tool usage as well as the preceding chat history. This compilation of data equips the LLM with the relevant information needed for informed decision-making in subsequent interactions.

The fourth element of the system prompt addresses traffic regulations and other essential considerations that the LLM must take into account. This includes adherence to specific traffic laws and the integration of vital factors that influence the decision-making process when utilizing the tools. It ensures that the LLM's actions are not only optimized for traffic flow but also compliant with legal and safety standards. The final component of the system prompt outlines the expected output format. By defining the output format, the system facilitates seamless communication between the LLM and the traffic control tools, ensuring that the decisions are implemented effectively and efficiently.

To further refine the decision-making process, each tool within the LA-Light system is accompanied by its own tailored prompt. These prompts are composed of four parts, enhancing the clarity and precision with which the tools can be operated. For instance, Table~\ref{tab_example_tool_prompt} shows the prompt structure for the tool \textit{Get\_Intersection\_Layout}. The first part provides a detailed description of the tool's capabilities. The second part outlines the necessary inputs required by the tool and the third part describes the output generated by the tool. Finally, a practical example is provided in Table~\ref{tab_example_tool_prompt} to illustrate how the tool is used and the example of the output.

\section{Experiments} \label{sec_experiment}

\subsection{Experiment Setting}

The experiments were carried out using the Simulation of Urban MObility (SUMO) \cite{lopez2018microscopic}, a widely recognized open source traffic simulator. To accurately capture traffic dynamics at intersections, we used virtual lane area detectors in the simulation to collect data such as vehicle count and queue length for each lane. Due to the constraints of simulated camera resolution, the scope of data collection was limited to a maximum of 150 meters from the intersection. This limitation was imposed to reflect realistic urban traffic monitoring conditions, although it is acknowledged that this may truncate actual queue lengths exceeding this distance.

In configuring the traffic signals, we adhered to common urban signaling sequences: a green light phase, followed by a 3-second yellow light, and then a red light phase. We set the parameters to match the realistic urban traffic flow, with a maximum speed limit of 13.9 m/s (i.e., 50 km/h). The minimum distance between vehicles was kept at 2.5 meters, consistent with safe driving distances in city environments. Vehicle speeds were modeled with a Gaussian distribution, having a mean of 10 m/s and a standard deviation of $\sqrt{3}$ m/s, to account for the variability in driver behavior. For the purposes of this study, we employed the GPT-4 Turbo model without any additional fine-tuning specific to TSC tasks. This choice was made to evaluate the model's out-of-the-box capabilities in managing complex traffic scenarios.

\subsection{Datasets and Scenarios}

\begin{figure*}[!ht]
    \centering
    \subfloat[]{\includegraphics[width=0.49\linewidth]{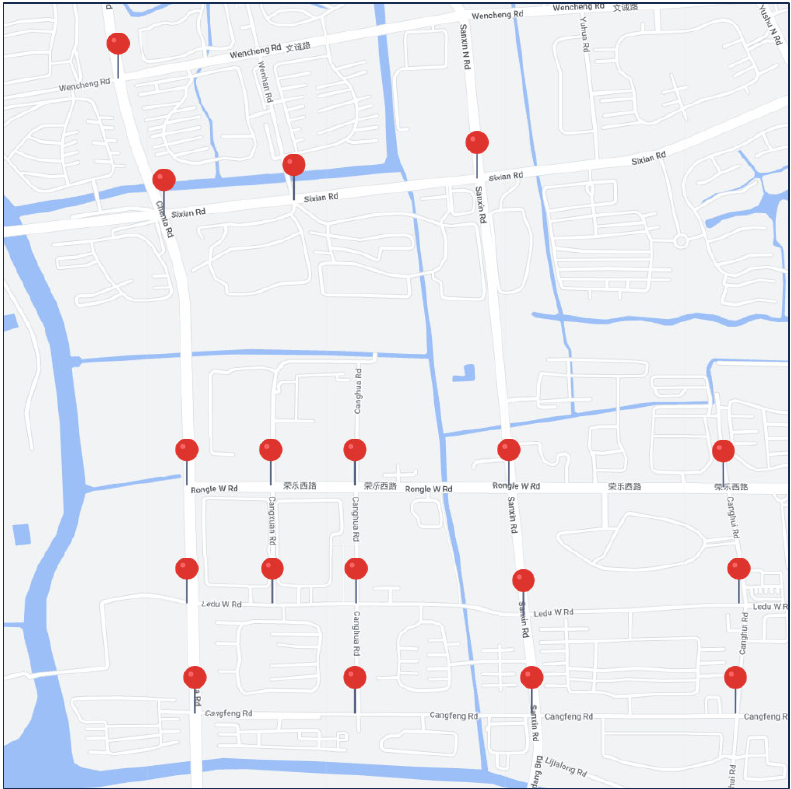}
    \label{fig_google_map}}
    \hfil
    \subfloat[]{\includegraphics[width=0.49\linewidth]{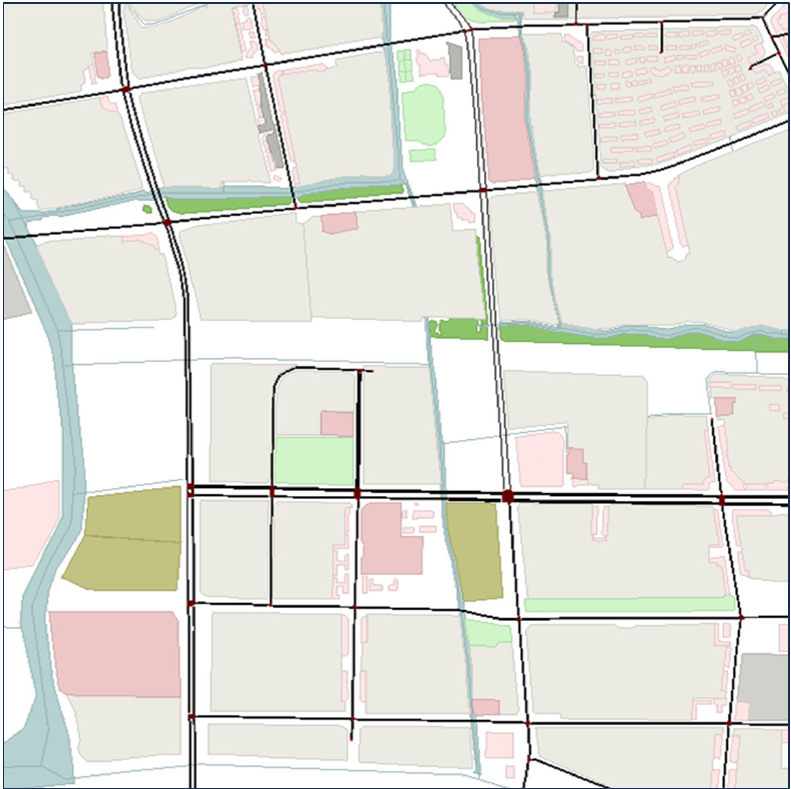}
    \label{fig_sumo_map}}
    \caption{Traffic Network at Chenta Road, Songjiang District, Shanghai. (a) on Google Maps. (b) in the SUMO simulator.}
    \label{fig_dataset_map}
\end{figure*}

Our experiment utilizes both synthetic and real-world datasets to evaluate the performance of the proposed LA-Light in traffic signal control. The synthetic dataset includes scenarios of isolated intersections with varying layouts: a three-way intersection and a four-way intersection, both featuring three lanes per approach. For real-world data, we focus on the urban road network surrounding Chenta Road in the Songjiang District of Shanghai, a region known for its heavy traffic congestion due to high-density construction and commercial activities. The network, illustrated in Fig.~\ref{fig_dataset_map}, encompasses 18 intersections, comprising a combination of twelve four-way and six three-way intersections. To collect traffic flow data, we analyzed video surveillance from these intersections on 30 July 2021. We recorded the number of vehicles per minute, which was then utilized to recreate the traffic scenarios in the SUMO platform.

\begin{figure*}[!ht]
    \centering
    \subfloat[]{\includegraphics[width=0.31\linewidth]{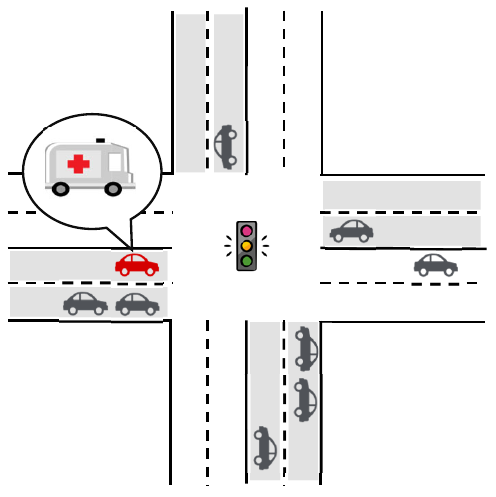}
    \label{fig_emv_scenario}}
    \hfil
    \subfloat[]{\includegraphics[width=0.31\linewidth]{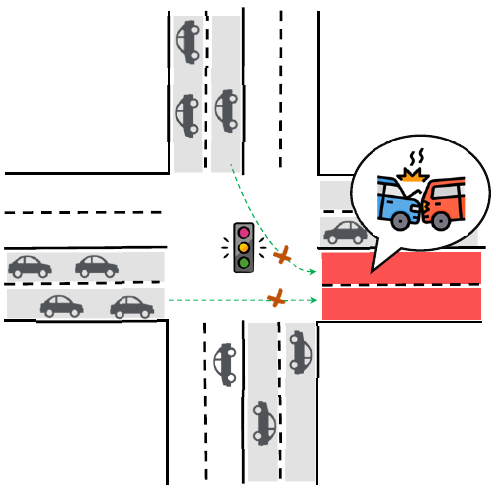}
    \label{fig_rbi_scenario}}
    \hfil
    \subfloat[]{\includegraphics[width=0.31\linewidth]{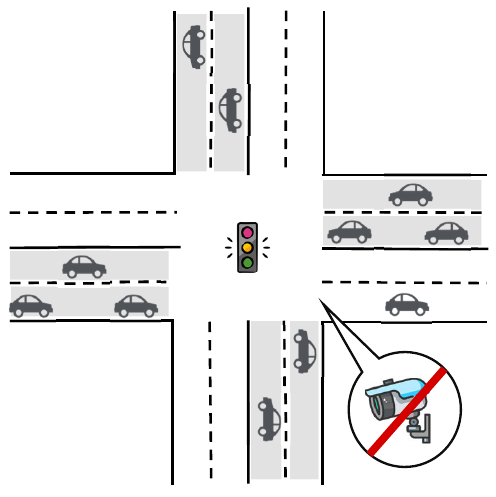}
    \label{fig_so_scenario}}
    \caption{Illustration of the three test scenarios. (a) Emergency Vehicle (EMV) Scenario, where ambulances are integrated into traffic flow; (b) Roadblock Incident (RBI) Scenario, depicting temporary road closures due to accidents or other events; (c) Sensor Outage (SO) Scenario, demonstrating the effects of sensor failures on traffic data accuracy.}
    \label{fig_scenarios}
\end{figure*}

To ensure a comprehensive evaluation of LA-Light's performance in complex urban traffic situations, we designed three specific test scenarios for each road network, as illustrated in Fig.~\ref{fig_scenarios}. The first scenario, depicted in Fig.~\ref{fig_emv_scenario}, is the \textbf{Emergency Vehicle (EMV) Scenario}. In this scenario, emergency vehicles, such as ambulances, are introduced into the normal traffic flow, making up $1\%$ of the overall traffic volume. These vehicles are assigned random origins and destinations to test the system's capability to prioritize them effectively. The second scenario, presented in Fig.~\ref{fig_rbi_scenario}, is the \textbf{Roadblock Incident (RBI) Scenario}. This scenario mimics the dynamic nature of urban traffic by introducing random roadblocks, which stand in for unexpected incidents like traffic accidents. These roadblocks occur for $10\%$ of the total simulation time, temporarily closing off affected lanes and testing the system's responsiveness to such events. The final scenario, shown in Fig.~\ref{fig_so_scenario}, is the \textbf{Sensor Outage (SO) Scenario}. This scenario simulates sensor reliability challenges by introducing a $10\%$ chance of sensor failure at any moment during the simulation. Such a failure results in the complete loss of vehicle detection data for that direction, challenging the system's ability to maintain efficient traffic control despite missing information.

\subsection{Metrics}

To evaluate the effectiveness of TSC strategies, this study utilizes a dual-perspective approach to metrics. We measure the Average Travel Time (ATT), which is the time taken by vehicles to travel from their origin to their destination. ATT is a critical metric for assessing traffic flow efficiency within the network. Alongside ATT, we examine the Average Waiting Time (AWT). AWT measures the average time vehicles spend traveling at speeds below 0.1 m/s, typically while waiting for the green signal, thus providing a direct measure of intersection delay.

To address the urgency of emergency services, we incorporate specific metrics for emergency vehicles. The Average Emergency Travel Time (AETT) and the Average Emergency Waiting Time (AEWT) are calculated separately to underscore the TSC's impact on prioritized vehicle movement. These metrics are essential for comparing TSC strategies that can adapt in emergency situations, an integral aspect of proficient urban traffic management.

\subsection{Compared Methods}

To assess the efficacy of the LA-Light model, we benchmarked it against a range of established TSC strategies. This benchmarking includes traditional transportation approaches as well as algorithms based on RL. We also evaluated the effectiveness of an LLM when used directly for decision-making, without the aid of Chain-of-Thought reasoning \cite{chu2023survey} and integration with existing TSC algorithms. The traditional traffic control methods are as follows:

\begin{itemize}
    \item \textbf{Webster} \cite{koonce2008traffic}: This approach involves calculating the optimal signal cycle lengths and phase splits based on current traffic volumes and signal phase sequences. For this study, we implemented a real-time version of Webster's method, which dynamically adjusts the traffic signals in response to actual traffic conditions.
    \item \textbf{SOTL} \cite{cools2013self}: The SOTL method evaluates the maximum queue length in the lanes associated with the current and subsequent signal phases. If the queue for the current phase is long, the green signal is extended; otherwise, the system triggers a shift to the next phase.
    \item \textbf{Maxpressure} \cite{varaiya2013max}: This advanced transportation method for traffic light control aims to reduce congestion in the lanes that exhibit the highest pressure. This pressure is quantified by the difference in queue lengths between upstream and downstream lanes. The method focuses on alleviating congestion where it is most needed.
\end{itemize}

For the RL-based models, we include:

\begin{itemize}
    \item \textbf{IntelliLight} \cite{wei2018intellilight}: This model leverages a state representation enriched with lane-specific details, including vehicle count and waiting time, thus providing a nuanced view of traffic conditions beyond simple queue length. The reward function of IntelliLight accounts for the frequency of signal changes, and it introduces a phase-gate model to mitigate the challenge of skewed phase data distribution.
    \item \textbf{PressLight} \cite{wei2019presslight}: An extension of the Maxpressure algorithm, PressLight combines deep reinforcement learning with the pressure optimization concept to dynamically adjust signals at intersections, aiming to maintain optimal flow.
    \item \textbf{AttendLight} \cite{oroojlooy2020attendlight}: Implementing an attention mechanism, AttendLight builds a set of observational features that inform the probability of phase changes, thus refining the signal timing optimization process.
    \item \textbf{UniTSA} \cite{wang2023unitsa}: Introducing a sophisticated intersection representation known as the junction matrix, UniTSA also brings five novel traffic state augmentation methods tailored to enhance signal control system performance.
\end{itemize}

Additionally, this study contrasts the proposed LA-Light system with a baseline LLM-based TSC method, herein referred to as \textbf{Vanilla-LLM}, which utilizes an LLM directly for TSC decision-making. Unlike LA-Light, the Vanilla-LLM approach does not incorporate Chain-of-Thought reasoning or supplemental decision-support tools. It relies solely on the model's inherent capabilities to interpret traffic data and make determinations based on the real-time traffic conditions. Key differences between the LA-Light framework and the Vanilla-LLM method are delineated in Table~\ref{tab:comparison}.

\begin{table}[!ht]
\centering
\caption{Comparison of LA-Light and Vanilla-LLM}
\label{tab:comparison}
\begin{tabular}{llcc}
\toprule
& \textbf{Tool/Component} & \textbf{Vanilla-LLM} & \textbf{LA-Light} \\
\midrule
\multicolumn{4}{l}{\textbf{Perception Tools (Static)}} \\
\cmidrule{2-4}
& Get\_Intersection\_Layout & $\times$ & $\checkmark$ \\
& Get\_Signal\_Phase\_Structure & $\times$ & $\checkmark$ \\
\midrule
\multicolumn{4}{l}{\textbf{Perception Tools (Dynamic)}} \\
\cmidrule{2-4}
& Get\_Occupancy & $\times$ & $\checkmark$ \\
& Get\_Queue\_Length & $\times$ & $\checkmark$ \\
& Get\_Phase\_ID & $\times$ & $\checkmark$ \\
& Get\_Junction\_Situation & $\times$ & $\checkmark$ \\
\midrule
\multicolumn{4}{l}{\textbf{Decision Tools (Support)}} \\
\cmidrule{2-4}
& Get\_Auxiliary\_Decision & $\times$ & $\checkmark$ \\
& Get\_Available\_Actions & $\times$ & $\checkmark$ \\
\midrule
\multicolumn{4}{l}{\textbf{Decision Tools (Evaluate)}} \\
\cmidrule{2-4}
& Evaluate\_Action\_Feasibility & $\times$ & $\checkmark$ \\
& Justify\_Decision\_Logic & $\times$ & $\checkmark$ \\
\midrule
\multicolumn{4}{l}{\textbf{Additional Capabilities}} \\
\cmidrule{2-4}
& Chain-of-Thought Reasoning & $\times$ & $\checkmark$ \\
& Chat History Analysis  & $\checkmark$ & $\checkmark$ \\
& Explains Recommendations & $\checkmark$ & $\checkmark$ \\
\bottomrule
\end{tabular}
\end{table}

\begin{table*}[!htbp]
\centering
\caption{Performance comparison under the Emergency Vehicle (EMV) Scenario. The best, second-best results are highlighted through \textbf{bold}, and \underline{underlining}, respectively.}
\label{tab_emv_scenario}
\resizebox{\textwidth}{!}{%
\begin{tabular}{ccccccccccccccc}
\hline
\multirow{2}{*}{Method} & \multicolumn{4}{c}{\textbf{3-Way INT}} &  & \multicolumn{4}{c}{\textbf{4 Way INT}} &  & \multicolumn{4}{c}{\textbf{Shanghai}} \\ \cline{2-5} \cline{7-10} \cline{12-15} 
 & ATT & AWT & AETT & AEWT &  & ATT & AWT & AETT & AEWT &  & ATT & AWT & AETT & AEWT \\ \hline
 & \multicolumn{14}{c}{\textbf{Traditional Transportation Approaches}} \\
Webster & 71.708 & 34.443 & 89.738 & 45.730 &  & 122.135 & 64.074 & 139.550 & 86.188 &  & 538.614 & 152.957 & 504.267 & 155.110 \\
SOTL & 68.275 & 31.100 & 77.475 & 35.954 &  & 104.569 & 53.364 & 100.087 & 53.975 &  & 498.995 & 121.054 & 453.162 & 105.125 \\
Maxpressure & 63.860 & 29.494 & 83.728 & 39.859 &  & 103.598 & 51.415 & 109.824 & 58.344 &  & 461.965 & 96.820 & 438.092 & 102.515 \\ \hline
 & \multicolumn{14}{c}{\textbf{RL-based Methods}} \\
IntelliLight & \underline{62.798} & 25.584 & 68.229 & 30.549 &  & 78.466 & 35.885 & 94.935 & 55.545 &  & 461.902 & 84.781 & 441.366 & 65.430 \\
PressLight & 67.736 & \textbf{20.609} & 81.275 & 36.411 &  & 71.143 & \underline{29.722} & 72.332 & 33.075 &  & \underline{407.352} & \textbf{78.020} & 423.658 & 109.210 \\
AttendLight & \textbf{61.750} & 26.796 & 77.926 & 34.229 &  & 74.538 & 32.341 & 88.569 & 46.743 &  & 429.452 & 81.017 & 432.419 & 116.944 \\
UniTSA & 67.848 & \underline{21.424} & 83.236 & 38.073 &  & \textbf{64.032} & \textbf{27.224} & 73.941 & 40.228 &  & \textbf{398.374} & \underline{79.426} & 403.642 & 98.593 \\ \hline
 & \multicolumn{14}{c}{\textbf{LLM-based Methods}} \\
Vanilla-LLM & 77.423 & 36.609 & \textbf{46.970} & \textbf{9.862} &  & 87.738 & 48.167 & \underline{49.324} & \underline{11.010} &  & 493.577 & 109.182 & \underline{391.836} & \underline{35.246} \\
LA-Light & 63.853 & 22.516 & \underline{48.824} & \underline{11.251} &  & \underline{69.965} & 31.457 & \textbf{47.808} & \textbf{10.435} &  & 411.826 & 82.802 & \textbf{371.997} & \textbf{17.476} \\ \hline
\end{tabular}}
\end{table*}

\begin{table*}[!htbp]
\centering
\caption{Performance comparison under the Roadblock Incident (RBI) Scenario.}
\label{tab_rbi_scenario}
\resizebox{\textwidth}{!}{%
\begin{tabular}{ccccccccccccccc}
\hline
\multirow{2}{*}{Method} & \multicolumn{4}{c}{\textbf{3-Way INT}} &  & \multicolumn{4}{c}{\textbf{4 Way INT}} &  & \multicolumn{4}{c}{\textbf{Shanghai}} \\ \cline{2-5} \cline{7-10} \cline{12-15} 
 & ATT & AWT & AETT & AEWT &  & ATT & AWT & AETT & AEWT &  & ATT & AWT & AETT & AEWT \\ \hline
 & \multicolumn{14}{c}{\textbf{Traditional Transportation Approaches}} \\
Webster & 83.019 & 38.768 & 102.043 & 59.879 &  & 132.069 & 80.311 & 138.364 & 82.314 &  & 596.965 & 282.545 & 674.380 & 177.341 \\
SOTL & 76.123 & 32.946 & 91.547 & 50.898 &  & 115.774 & 61.089 & 114.064 & 61.606 &  & 593.314 & 244.136 & 539.789 & 207.316 \\
Maxpressure & 73.811 & 32.910 & 96.754 & 58.398 &  & 115.573 & 64.153 & 102.236 & 54.278 &  & 512.578 & 118.984 & 477.416 & 92.406 \\ \hline
 & \multicolumn{14}{c}{\textbf{RL-based Methods}} \\
IntelliLight & \underline{68.007} & \underline{29.178} & 71.237 & 38.331 &  & 80.973 & 48.819 & 90.449 & 47.794 &  & 494.662 & 117.503 & 475.623 & 106.845 \\
PressLight & 76.883 & 33.909 & 97.470 & 53.384 &  & 83.594 & 37.166 & 84.441 & 32.813 &  & 487.390 & 106.350 & 453.702 & 106.651 \\
AttendLight & 69.418 & 29.849 & 92.717 & 54.235 &  & \underline{78.364} & \underline{35.330} & 76.735 & 26.958 &  & 493.844 & 121.402 & 477.328 & 131.610 \\
UniTSA & 79.760 & 43.367 & 83.354 & 49.510 &  & 81.650 & 38.604 & 78.828 & 38.693 &  & \underline{467.262} & \textbf{97.154} & \underline{410.960} & 70.844 \\ \hline
 & \multicolumn{14}{c}{\textbf{LLM-based Methods}} \\
Vanilla-LLM & 79.967 & 34.122 & \textbf{54.912} & \textbf{15.939} &  & 93.009 & 47.206 & \underline{73.451} & \underline{24.523} &  & 490.522 & 93.035 & 420.371 & \underline{51.674} \\
LA-Light & \textbf{66.510} & \textbf{27.208} & \underline{55.094} & \underline{20.458} &  & \textbf{71.982} & \textbf{33.266} & \textbf{64.808} & \textbf{18.900} &  & \textbf{435.698} & \underline{86.902} & \textbf{379.822} & \textbf{23.031} \\ \hline
\end{tabular}}
\end{table*}

\begin{table*}[!htbp]
\centering
\caption{Performance comparison under the Sensor Outage (SO) Scenario.}
\label{tab_so_scenario}
\resizebox{\textwidth}{!}{%
\begin{tabular}{ccccccccccccccc}
\hline
\multirow{2}{*}{Method} & \multicolumn{4}{c}{\textbf{3-Way INT}} &  & \multicolumn{4}{c}{\textbf{4 Way INT}} &  & \multicolumn{4}{c}{\textbf{Shanghai}} \\ \cline{2-5} \cline{7-10} \cline{12-15} 
 & ATT & AWT & AETT & AEWT &  & ATT & AWT & AETT & AEWT &  & ATT & AWT & AETT & AEWT \\ \hline
 & \multicolumn{14}{c}{\textbf{Traditional Transportation Approaches}} \\
Webster & 78.449 & 33.250 & 82.845 & 45.740 &  & 124.008 & 67.745 & 134.664 & 80.372 &  & 548.485 & 159.720 & 582.200 & 202.231 \\
SOTL & 71.969 & 31.543 & 65.527 & 47.775 &  & 109.181 & 55.314 & 107.564 & 56.215 &  & 505.916 & 151.816 & 487.549 & 158.348 \\
Maxpressure & \underline{69.076} & 28.555 & 68.010 & 39.950 &  & 107.863 & 62.312 & 127.463 & 82.852 &  & 538.497 & 128.103 & 551.391 & 174.590 \\ \hline
 & \multicolumn{14}{c}{\textbf{RL-based Methods}} \\
IntelliLight & 77.417 & 38.152 & 83.909 & 41.413 &  & 95.080 & 54.047 & 111.095 & 70.785 &  & 517.549 & 235.881 & 594.908 & 318.346 \\
PressLight & 80.850 & 35.989 & 79.727 & 36.406 &  & 91.357 & 48.579 & 90.753 & 48.266 &  & 534.881 & 227.570 & 618.964 & 329.579 \\
AttendLight & 73.108 & \underline{29.473} & 77.385 & 36.563 &  & 86.714 & 41.764 & 99.366 & 55.019 &  & 557.932 & 192.267 & 625.205 & 282.224 \\
UniTSA & 83.429 & 50.217 & 89.839 & 47.719 &  & \underline{81.370} & \underline{39.792} & 82.158 & 41.203 &  & \underline{474.909} & \textbf{103.004} & 497.134 & 124.262 \\ \hline
 & \multicolumn{14}{c}{\textbf{LLM-based Methods}} \\
Vanilla-LLM & 81.125 & 44.184 & \underline{49.412} & \underline{13.754} &  & 85.993 & 40.698 & \underline{56.215} & \underline{13.356} &  & 496.504 & 109.405 & \underline{407.128} & \underline{38.696} \\
LA-Light & \textbf{67.726} & \textbf{23.520} & \textbf{46.568} & \textbf{10.759} &  & \textbf{72.071} & \textbf{33.874} & \textbf{47.204} & \textbf{9.741} &  & \textbf{438.408} & \underline{82.600} & \textbf{380.228} & \textbf{21.086} \\ \hline
\end{tabular}}
\end{table*}

\subsection{Performance Analysis}

In this section, we assess the performance of the proposed LA-Light framework alongside various benchmark methods across three distinct road maps, with each map featuring three unique scenarios. Table~\ref{tab_emv_scenario} presents the results under the EMV Scenario. LA-Light's comparative analysis against traditional traffic signal control methods, RL-based approaches, and other LLM-based methods demonstrates a comprehensive enhancement in traffic signal control efficiency for both regular and emergency vehicles. For example, in comparison with the Maxpressure approach, LA-Light achieves a $32.1\%$ reduction in ATT for the four-way intersection (4-Way INT) and a $10.8\%$ reduction for the Shanghai network. In terms of emergency vehicle efficiency, indicated by AETT, LA-Light shows a significant improvement, reducing AETT by $15.3\%$ in the Shanghai network compared to Maxpressure. This improvement is attributed to the integration of RL algorithms within LA-Light, which refines decision-making processes in standard traffic scenarios.

When compared with RL-based methods, LA-Light does not always surpass in ATT and AWT due to its prioritization of emergency vehicles, which can extend the wait for other vehicles. In contrast, RL-based methods typically do not account for emergency vehicle priority. Consequently, LA-Light achieves significantly better efficiency for emergency response. For example, while AttendLight achieves a $3.3\%$ lower ATT at a three-way intersection scenario (3-Way INT), LA-Light shows a remarkable $67.3\%$ reduction in AEWT for the same condition. This highlights LA-Light's capability to assimilate environmental observations and adjust to dynamic changes, emphasizing its robustness in urgent situations without specialized fine-tuning. Furthermore, when LA-Light is compared to another LLM-based method, Vanilla-LLM, it exhibits a notable improvement in ATT and AWT across all tested networks. Specifically, in the complex Shanghai network, LA-Light reduces ATT and AWT by $16.5\%$ and $24.2\%$, respectively, compared to Vanilla-LLM. The integration of existing TSC methods within LA-Light's decision-making framework likely contributes to its enhanced traffic efficiency, demonstrating the potential of LLM-assisted approaches in urban traffic management.

Table~\ref{tab_rbi_scenario} details the performance outcomes in the RBI Scenario, where LA-Light's adaptability to unforeseen traffic events is pronounced. Traditional transportation methods like Webster, SOTL, and Maxpressure, which depend on fixed algorithms, are less adept at adjusting to sudden changes such as those introduced by roadblocks. RL-based methods are more adaptable but are still constrained by their reliance on previously learned strategies, which may not be sufficiently flexible for drastic alterations in road capacity. LA-Light, leveraging the real-time processing abilities of its LLM, dynamically responds to these traffic alterations. In the 3-Way INT scenario, LA-Light shows a $2.2\%$ reduction in ATT and a $6.3\%$ reduction in AWT in comparison to IntelliLight, the most effective RL-based method. In the more intricate Shanghai network, LA-Light's performance is even more notable, with a $6.8\%$ improvement in ATT and a $11.3\%$ improvement in AWT over UniTSA, the best RL-based approach.

Furthermore, LA-Light's proficiency extends to emergency response metrics, such as AETT and AEWT. On the 4-way INT, LA-Light shows a $35.6\%$ betterment in AETT and a $74.5\%$ enhancement in AEWT compared to UniTSA. Against another LLM-based method, Vanilla-LLM, LA-Light underscores the value of not only LLM's decision-making capabilities but also the sophisticated integration of chain-of-thought reasoning with effective tool utilization. This is particularly evident in the Shanghai network, where LA-Light achieves a $16.6\%$ reduction in ATT and a remarkable $51.4\%$ improvement in AEWT compared to Vanilla-LLM.

Finally, Table~\ref{tab_so_scenario} shows the performance under SO Scenario. Similar to the conclusion under RBI Scenario, the result demonstrates LA-Light's capability to effectively manage traffic even with sensor failures, a rare but critical challenge. For example, in the complex Shanghai network, LA-Light significantly reduces ATT and AWT by $20.0\%$ and $35.9\%$, respectively, compared to the Maxpressure method. While RL-based methods exhibit a degree of adaptability, they struggle in the absence of sensor data. Our method, LA-Light, addresses this shortcoming by utilizing common sense reasoning and the tools at hand. Compared to UniTSA, the top-performing RL-based method in this scenario, LA-Light achieves a $7.7\%$ improvement in ATT and a $20.4\%$ reduction in AWT. Moreover, LA-Light's performance excels against another LLM-based method, Vanilla-LLM, with an $11.7\%$ betterment in ATT and a $24.8\%$ enhancement in AWT, highlighting the efficiency of its decision-making process in scenarios with incomplete data, and confirming its robustness as a reliable traffic management solution.

The comparative analysis of LA-Light's performance in varying scenarios underscores its reliable effectiveness amidst environmental uncertainties. Notably, within the Shanghai network, the shift from the EMV to the SO scenario resulted in a modest increase $6.9\%$ in ATT and a $0.2\%$ in AWT, demonstrating LA-Light's commendable stability. This performance is markedly superior to RL-based methods, such as UniTSA, which exhibited a significant performance drop, $16.1\%$ in ATT and $23.3\%$ in AWT, under the same conditions. Further, LA-Light's emergency response metrics, specifically AETT and AEWT, remain the best among all benchmarks in all three scenarios. This consistency confirms the resilience of the LA-Light framework, which leverages LLMs to ensure minimal performance decline even in less common situations. These results emphasize LA-Light's capability to deliver dependable traffic signal control in diverse and complex urban environments.

\begin{figure*}[!htbp]
    \centering
    \includegraphics[width=0.99\linewidth]{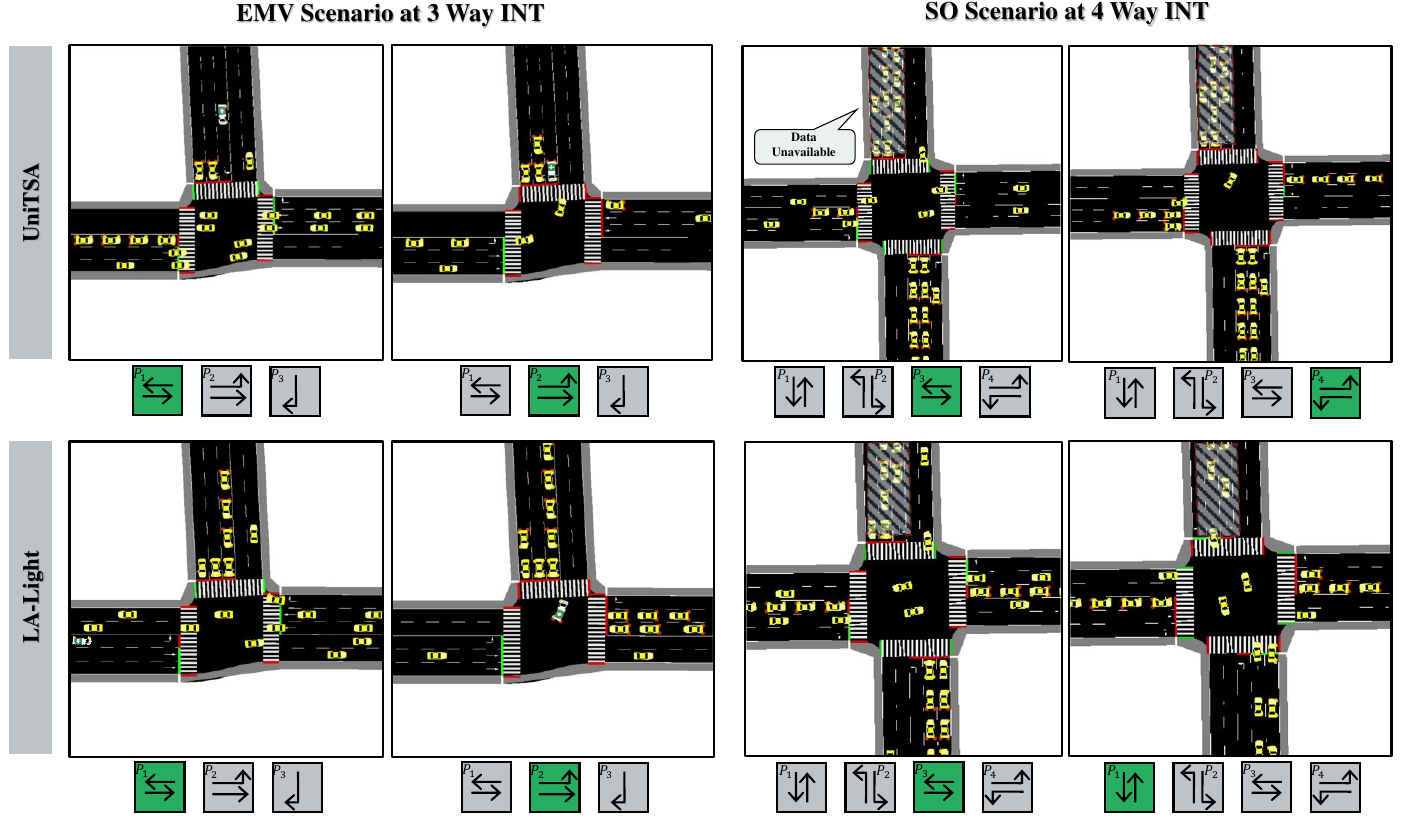}
    \caption{Comparative analysis of UniTSA and LA-Light strategies in handling EMV and SO scenarios at synthetic intersections. The top row illustrates the decision and subsequent traffic flow impact of the UniTSA method, while the bottom row demonstrates the same for the LA-Light method. The sequence of actions and their effects on traffic dynamics are depicted, with the left side of each panel displaying the intersection conditions considered by each agent, and the right side showing the implemented actions and resultant traffic flow alterations at 3 Way INT and 4 Way INT.}
    \label{fig:int_case}
\end{figure*}

\begin{figure*}[!htbp]
    \centering
    \includegraphics[width=0.99\linewidth]{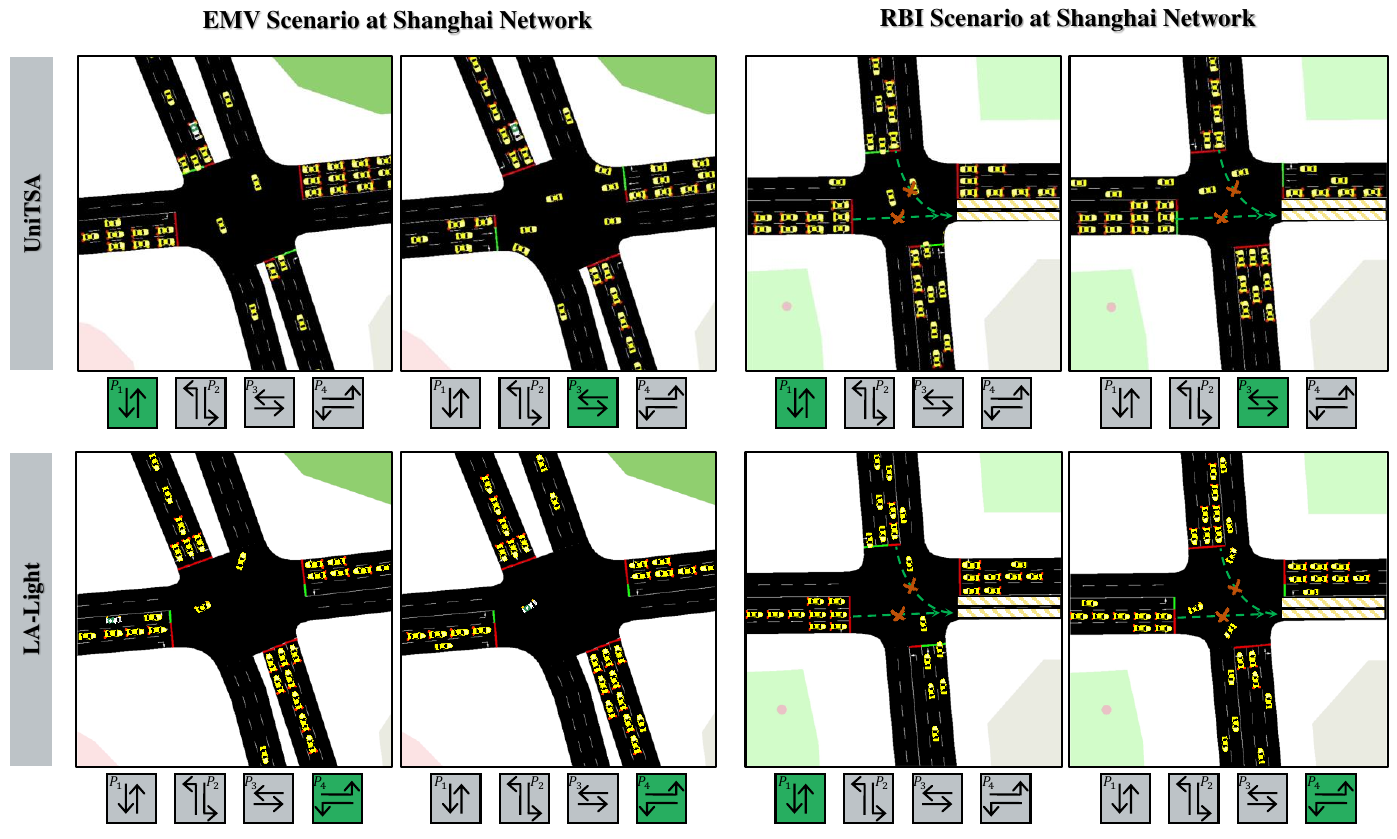}
    \caption{Comparative analysis of UniTSA and LA-Light strategies in handling EMV and RBI scenarios at the Shanghai network.}
    \label{fig:shanghai_case}
\end{figure*}

\begin{figure*}[!htbp]
    \centering
    \includegraphics[width=0.9\linewidth]{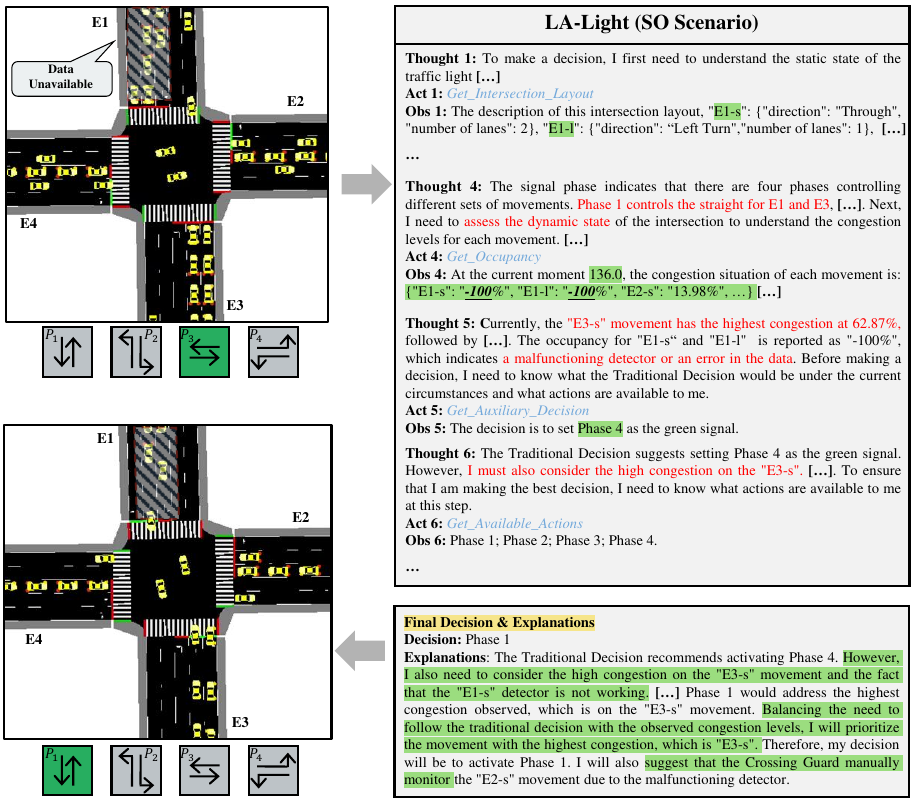}
    \caption{Decision-making process and explanation of LA-Light during the SO Scenario.}
    \label{fig:so_scenario_steps}
\end{figure*}

\begin{figure*}[!htbp]
    \centering
    \includegraphics[width=0.9\linewidth]{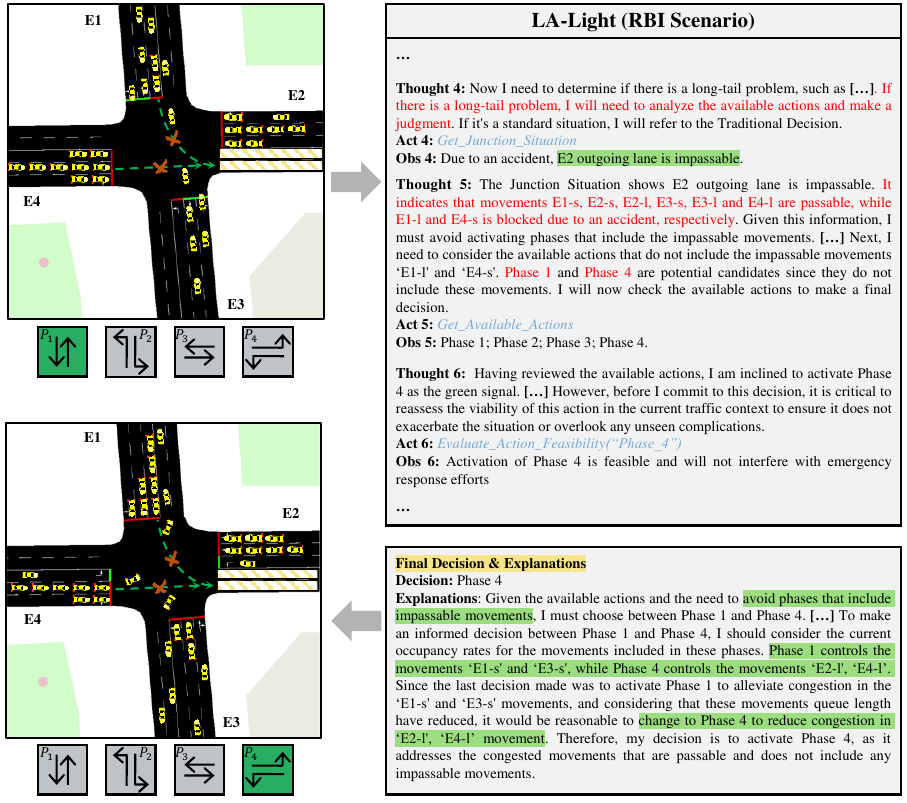}
    \caption{Decision-making process and explanation of LA-Light during the RBI Scenario.}
    \label{fig:rbi_scenario_steps}
\end{figure*}

\subsection{Case Study Insights}

In this section, we conduct a comprehensive analysis of the decision-making processes utilized by the LA-Light across a variety of traffic conditions. We first compare the decision of our method with that of the UniTSA in different traffic scenarios. These results are depicted in Fig.~\ref{fig:int_case} for the synthetic dataset and Fig.~\ref{fig:shanghai_case} for scenarios in Shanghai. In the synthetic dataset, as depicted in Fig.~\ref{fig:int_case}, the LA-Light framework demonstrates enhanced adaptability. For example, in the EMV Scenario at the 3 Way INT, UniTSA changes the signal from phase $P_{1}$ to $P_{2}$, giving priority to the larger volume of vehicles turning left from the west side (W-l). However, this action inadvertently causes a delay for the emergency vehicle. In contrast, LA-Light, utilizing the LLM's nuanced understanding of the scenario, changes the green phase to $P_{2}$, which, despite the queue forming on the north side (N-l), allows the emergency vehicle to pass without delay. Similarly, in the SO Scenario at the 4 Way INT, UniTSA fails to receive accurate data from the north side due to a damaged sensor. As a result, it mistakenly extends the green phase for $P_{4}$, which worsens the congestion on the north side. LA-Light, on the other hand, identifies the faulty sensor data and integrates this information with the real-time traffic conditions on the south side. It infers that congestion is increasing on the north side and accordingly adjusts the signal to green phase $P_{1}$, which is for the north-south through traffic, effectively reducing the congestion at the intersection.

The real-world scenarios presented in Fig.~\ref{fig:shanghai_case} further highlight the effectiveness of the LA-Light system in navigating the complexities of urban traffic networks. In the EMV Scenario within the Shanghai network, LA-Light aptly prolongs the current green phase $P_{4}$, allowing the emergency vehicle to pass swiftly. In contrast, UniTSA does not give precedence to the emergency vehicle, opting instead to clear lanes with higher vehicle accumulation, thereby neglecting the urgency of emergency response. In the RBI Scenario, where an obstruction is present on the east exit road, LA-Light successfully infers the larger traffic impact and redirects the green phase towards lanes that are not affected, preventing further congestion. UniTSA, however, defaults to activating the green phase for $P_{3}$, which is rendered ineffective as the blockage hinders northbound traffic from proceeding, leading to a suboptimal use of the green phase. These case studies demonstrate that LA-Light can dynamically adapt to diverse and unpredictable environments, showcasing its potential for zero-shot adaptation in real-time traffic management.

In the following analysis, we explore the decision-making process of LA-Light in response to various urban traffic scenarios. Fig.~\ref{fig:so_scenario_steps} illustrates the steps taken under the SO Scenario. LA-Light begins its evaluation by analyzing both the static and dynamic aspects of the traffic intersection. These include the intersection's physical configuration, preset signal phase timings, and real-time traffic density for each lane and direction. During this initial phase, LA-Light identifies a discrepancy: the sensor on the ``E1-s'' (E1 straight) approach reports an occupancy of $-100\%$, clearly signaling a fault, as negative occupancy is not feasible. Simultaneously, it notes significant congestion on the ``E3-s'' direction, with an occupancy of $62.87\%$, and the ``E2-l'' direction, with $54.00\%$ occupancy. Facing incomplete data, LA-Light uses an auxiliary decision-support tool for a reference solution. This tool suggests prioritizing phase $P_{4}$, managing left turns from E2 and E4 approaches, aiming to reduce queue lengths, especially for the congested ``E2-l'' movement.

However, LA-Light applies more complex reasoning. It considers the heavy congestion on the ``E3-s'' direction and the inoperative ``E1-s'' sensor, requiring a strategic response. The faulty sensor makes it impossible to estimate the queue on the ``E1-s'' approach accurately, which may be significant. LA-Light decides to deviate from UniTSA's advice, opting for phase $P_{1}$ to alleviate traffic from the ``E1-s'' and ``E3-s'' directions, likely facing higher demand. Additionally, LA-Light recommends a prompt on-site inspection of the ``E1-s'' approach by traffic personnel to counter unreliable sensor data. This human intervention ensures the resilience of the decision-making process, even when automated systems are compromised.

Fig.~\ref{fig:rbi_scenario_steps} illustrates LA-Light's decision-making process in the RBI Scenario, starting with gathering crucial data from the intersection as done in the SO Scenario depicted in Fig.~\ref{fig:so_scenario_steps}. The ``get junction situation'' function is then used, revealing a blockage on the E2 outgoing edge due to a vehicular accident. After that, LA-Light performs a detailed evaluation of the intersection's condition, determining that the ``E1-l'' and ``E4-s'' movements, associated with $P_{2}$ and $P_{3}$ phases of the traffic signals, are the most affected by the accident.

Having pinpointed the affected phases, LA-Light employs the tool to list all potential signal phases, then eliminates those hindered by the incident. This process leaves $P_{1}$ and $P_{4}$ as the feasible choices. In choosing between them, LA-Light examines the current congestion at the intersection. The decision-making is strategically grounded: the previous decision involved setting phase $P_{1}$ to a green signal to reduce congestion in the ``E1-s'' and ``E3-s'' directions. This move has effectively shortened queue lengths according to the latest data. Thus, to continue efficient traffic management and to alleviate congestion elsewhere, LA-Light opts for $P_{4}$. This phase controls the ``E2-l'' and ``E4-l'' movements, which, though congested, are clear of obstructions and can therefore benefit from a longer green signal. The selection of $P_{4}$ is deliberate, in line with LA-Light's goal to manage congestion proactively while avoiding the areas affected by the accident. This decision showcases LA-Light's capacity to adjust dynamically to real-time traffic situations, using its computational capabilities to maintain traffic flow as smoothly as possible despite unexpected challenges. Additionally, the rationale behind the decisions made in both the SO and RBI scenarios is also clarified in Fig.~\ref{fig:so_scenario_steps} and Fig.~\ref{fig:rbi_scenario_steps}. This transparency in the decision-making process enhances the reliability and trustworthiness of the signal control system.

\section{Conclusion} \label{sec_conclude}

In this work, we have presented the LA-Light framework, which incorporates LLMs to improve decision-making in the dynamic and complex environment of urban traffic management. By combining the sophisticated reasoning abilities of LLMs with established TSC methods and real-time data collection tools, we have established a new approach to traffic signal control. Our comprehensive evaluation of LA-Light, conducted across three distinct traffic networks and nine unique scenarios, has shown its effectiveness without necessitating further training. Compared to traditional methods, LA-Light has consistently achieved reductions in ATT and AWT. Additionally, it has demonstrated improvements in metrics for emergency response vehicles, such as AETT and AEWT. Analysis of the decision-making process in various contexts revealed that LA-Light excels not only in operational performance but also in decision-making clarity, courtesy of the LLMs' explanatory capabilities. LA-Light adeptly identifies and utilizes the most suitable tools for a given traffic situation, providing clear insights into its decision-making rationale.

While the LA-Light framework marks a significant step forward, it does have areas that require further refinement. The framework's current dependency on frequent interactions with the LLM for decision-making introduces a delay that could impact the promptness of traffic signal adjustments. Moreover, the framework's reliance on textual descriptions to depict traffic scenarios may not encompass all the details needed for the most effective decision-making, pointing to the potential benefits of a more direct, image-based approach that can interpret traffic conditions from visual data. Future work will aim to address these issues by refining the interaction process to expedite response times and by incorporating vision-based models capable of directly processing visual information. These enhancements are expected to improve the framework's proficiency in managing the complexities of real-world traffic systems with increased speed and less reliance on textual descriptions.

\section{Acknowledgements}

This work was supported in part by Shanghai Pujiang Program (21PJD092), the National Key R$\&$D Program of China with grant No. 2018YFB1800800, the Basic Research Project No. HZQB-KCZYZ-2021067 of Hetao Shenzhen-HK S$\&$T Cooperation Zone. 

\bibliographystyle{ieeetr}
\bibliography{references}

\begin{thebibliography}{10}

\bibitem{sweet2011does}
M.~Sweet, ``Does traffic congestion slow the economy?,'' {\em Journal of Planning Literature}, vol.~26, no.~4, pp.~391--404, 2011.

\bibitem{zhao2011computational}
D.~Zhao, Y.~Dai, and Z.~Zhang, ``Computational intelligence in urban traffic signal control: A survey,'' {\em IEEE Transactions on Systems, Man, and Cybernetics, Part C (Applications and Reviews)}, vol.~42, no.~4, pp.~485--494, 2011.

\bibitem{koonce2008traffic}
P.~Koonce and L.~Rodegerdts, ``Traffic signal timing manual.,'' tech. rep., United States. Federal Highway Administration, 2008.

\bibitem{cools2013self}
S.-B. Cools, C.~Gershenson, and B.~D’Hooghe, ``Self-organizing traffic lights: A realistic simulation,'' {\em Advances in applied self-organizing systems}, pp.~45--55, 2013.

\bibitem{qadri2020state}
S.~S. S.~M. Qadri, M.~A. G{\"o}k{\c{c}}e, and E.~{\"O}ner, ``State-of-art review of traffic signal control methods: challenges and opportunities,'' {\em European transport research review}, vol.~12, pp.~1--23, 2020.

\bibitem{wei2019survey}
H.~Wei, G.~Zheng, V.~Gayah, and Z.~Li, ``A survey on traffic signal control methods,'' {\em arXiv preprint arXiv:1904.08117}, 2019.

\bibitem{vardhan2021rare}
H.~Vardhan and J.~Sztipanovits, ``Rare event failure test case generation in learning-enabled-controllers,'' in {\em 6th International Conference on Machine Learning Technologies}, pp.~34--40, 2021.

\bibitem{martinez2011survey}
F.~J. Martinez, C.~K. Toh, J.-C. Cano, C.~T. Calafate, and P.~Manzoni, ``A survey and comparative study of simulators for vehicular ad hoc networks ({VANETs}),'' {\em Wireless Communications and Mobile Computing}, vol.~11, no.~7, pp.~813--828, 2011.

\bibitem{hunt1982scoot}
P.~Hunt, D.~Robertson, R.~Bretherton, and M.~C. Royle, ``The scoot on-line traffic signal optimisation technique,'' {\em Traffic Engineering \& Control}, vol.~23, no.~4, 1982.

\bibitem{lowrie1990scats}
P.~Lowrie, ``Scats-a traffic responsive method of controlling urban traffic,'' {\em Sales information brochure published by Roads \& Traffic Authority, Sydney, Australia}, 1990.

\bibitem{wu2023deep}
C.~Wu, I.~Kim, and Z.~Ma, ``Deep reinforcement learning based traffic signal control: A comparative analysis,'' {\em Procedia Computer Science}, vol.~220, pp.~275--282, 2023.

\bibitem{wei2019presslight}
H.~Wei, C.~Chen, G.~Zheng, K.~Wu, V.~Gayah, K.~Xu, and Z.~Li, ``Presslight: Learning max pressure control to coordinate traffic signals in arterial network,'' in {\em Proceedings of the 25th ACM SIGKDD International Conference on Knowledge Discovery \& Data Mining}, pp.~1290--1298, 2019.

\bibitem{zang2020metalight}
X.~Zang, H.~Yao, G.~Zheng, N.~Xu, K.~Xu, and Z.~Li, ``Metalight: Value-based meta-reinforcement learning for traffic signal control,'' in {\em Proceedings of the AAAI Conference on Artificial Intelligence}, vol.~34, pp.~1153--1160, 2020.

\bibitem{chen2020toward}
C.~Chen, H.~Wei, N.~Xu, G.~Zheng, M.~Yang, Y.~Xiong, K.~Xu, and Z.~Li, ``Toward a thousand lights: Decentralized deep reinforcement learning for large-scale traffic signal control,'' in {\em Proceedings of the AAAI Conference on Artificial Intelligence}, vol.~34, pp.~3414--3421, 2020.

\bibitem{pang2024scalable}
A.~Pang, M.~Wang, Y.~Chen, M.-O. Pun, and M.~Lepech, ``Scalable reinforcement learning framework for traffic signal control under communication delays,'' {\em IEEE Open Journal of Vehicular Technology}, 2024.

\bibitem{gu2024large}
H.~Gu, S.~Wang, X.~Ma, D.~Jia, G.~Mao, E.~G. Lim, and C.~P.~R. Wong, ``Large-scale traffic signal control using constrained network partition and adaptive deep reinforcement learning,'' {\em IEEE Transactions on Intelligent Transportation Systems}, pp.~1--14, 2024.

\bibitem{chu2019multi}
T.~Chu, J.~Wang, L.~Codec{\`a}, and Z.~Li, ``Multi-agent deep reinforcement learning for large-scale traffic signal control,'' {\em IEEE Transactions on Intelligent Transportation Systems}, vol.~21, no.~3, pp.~1086--1095, 2019.

\bibitem{wang2022adlight}
M.~Wang, Y.~Xu, X.~Xiong, Y.~Kan, C.~Xu, and M.-O. Pun, ``{ADLight}: A universal approach of traffic signal control with augmented data using reinforcement learning,'' in {\em Transportation Research Board (TRB) 102nd Annual Meeting}, 2023.

\bibitem{wang2023unitsa}
M.~Wang, X.~Xiong, Y.~Kan, C.~Xu, and M.-O. Pun, ``{UniTSA}: A universal reinforcement learning framework for v2x traffic signal control,'' {\em IEEE Transactions on Vehicular Technology}, pp.~1--16, 2024.

\bibitem{varaiya2013max}
P.~Varaiya, ``Max pressure control of a network of signalized intersections,'' {\em Transportation Research Part C: Emerging Technologies}, vol.~36, pp.~177--195, 2013.

\bibitem{oroojlooy2020attendlight}
A.~Oroojlooy, M.~Nazari, D.~Hajinezhad, and J.~Silva, ``Attendlight: Universal attention-based reinforcement learning model for traffic signal control,'' {\em Advances in Neural Information Processing Systems}, vol.~33, pp.~4079--4090, 2020.

\bibitem{jiang2024general}
H.~Jiang, Z.~Li, Z.~Li, L.~Bai, H.~Mao, W.~Ketter, and R.~Zhao, ``A general scenario-agnostic reinforcement learning for traffic signal control,'' {\em IEEE Transactions on Intelligent Transportation Systems}, pp.~1--15, 2024.

\bibitem{wei2018intellilight}
H.~Wei, G.~Zheng, H.~Yao, and Z.~Li, ``Intellilight: A reinforcement learning approach for intelligent traffic light control,'' in {\em Proceedings of the 24th ACM SIGKDD International Conference on Knowledge Discovery \& Data Mining}, pp.~2496--2505, 2018.

\bibitem{bouktif2023deep}
S.~Bouktif, A.~Cheniki, A.~Ouni, and H.~El-Sayed, ``Deep reinforcement learning for traffic signal control with consistent state and reward design approach,'' {\em Knowledge-Based Systems}, vol.~267, p.~110440, 2023.

\bibitem{floridi2020gpt}
L.~Floridi and M.~Chiriatti, ``{GPT}-3: Its nature, scope, limits, and consequences,'' {\em Minds and Machines}, vol.~30, pp.~681--694, 2020.

\bibitem{chatgpt2023}
OpenAI, ``Introducing {ChatGPT}.'' \url{https://openai.com/blog/chatgpt/}, 2023.

\bibitem{touvron2023llama}
H.~Touvron, T.~Lavril, G.~Izacard, X.~Martinet, M.-A. Lachaux, T.~Lacroix, B.~Rozi{\`e}re, N.~Goyal, E.~Hambro, F.~Azhar, {\em et~al.}, ``Llama: Open and efficient foundation language models,'' {\em arXiv preprint arXiv:2302.13971}, 2023.

\bibitem{touvron2023llama2}
H.~Touvron, L.~Martin, K.~Stone, P.~Albert, A.~Almahairi, Y.~Babaei, N.~Bashlykov, S.~Batra, P.~Bhargava, S.~Bhosale, {\em et~al.}, ``Llama 2: Open foundation and fine-tuned chat models,'' {\em arXiv preprint arXiv:2307.09288}, 2023.

\bibitem{vaswani2017attention}
A.~Vaswani, N.~Shazeer, N.~Parmar, J.~Uszkoreit, L.~Jones, A.~N. Gomez, {\L}.~Kaiser, and I.~Polosukhin, ``Attention is all you need,'' {\em Advances in neural information processing systems}, vol.~30, 2017.

\bibitem{ouyang2022training}
L.~Ouyang, J.~Wu, X.~Jiang, D.~Almeida, C.~Wainwright, P.~Mishkin, C.~Zhang, S.~Agarwal, K.~Slama, A.~Ray, {\em et~al.}, ``Training language models to follow instructions with human feedback,'' {\em Advances in Neural Information Processing Systems}, vol.~35, pp.~27730--27744, 2022.

\bibitem{wei2022chain}
J.~Wei, X.~Wang, D.~Schuurmans, M.~Bosma, F.~Xia, E.~Chi, Q.~V. Le, D.~Zhou, {\em et~al.}, ``Chain-of-thought prompting elicits reasoning in large language models,'' {\em Advances in Neural Information Processing Systems}, vol.~35, pp.~24824--24837, 2022.

\bibitem{yao2022react}
S.~Yao, J.~Zhao, D.~Yu, N.~Du, I.~Shafran, K.~Narasimhan, and Y.~Cao, ``React: Synergizing reasoning and acting in language models,'' {\em arXiv preprint arXiv:2210.03629}, 2022.

\bibitem{zhao2023survey}
W.~X. Zhao, K.~Zhou, J.~Li, T.~Tang, X.~Wang, Y.~Hou, Y.~Min, B.~Zhang, J.~Zhang, Z.~Dong, {\em et~al.}, ``A survey of large language models,'' {\em arXiv preprint arXiv:2303.18223}, 2023.

\bibitem{xi2023rise}
Z.~Xi, W.~Chen, X.~Guo, W.~He, Y.~Ding, B.~Hong, M.~Zhang, J.~Wang, S.~Jin, E.~Zhou, {\em et~al.}, ``The rise and potential of large language model based agents: A survey,'' {\em arXiv preprint arXiv:2309.07864}, 2023.

\bibitem{cui2024survey}
C.~Cui, Y.~Ma, X.~Cao, W.~Ye, Y.~Zhou, K.~Liang, J.~Chen, J.~Lu, Z.~Yang, K.-D. Liao, {\em et~al.}, ``A survey on multimodal large language models for autonomous driving,'' in {\em Proceedings of the IEEE/CVF Winter Conference on Applications of Computer Vision}, pp.~958--979, 2024.

\bibitem{cui2023receive}
C.~Cui, Y.~Ma, X.~Cao, W.~Ye, and Z.~Wang, ``Receive, reason, and react: Drive as you say with large language models in autonomous vehicles,'' {\em arXiv preprint arXiv:2310.08034}, 2023.

\bibitem{da2023open}
L.~Da, K.~Liou, T.~Chen, X.~Zhou, X.~Luo, Y.~Yang, and H.~Wei, ``{Open-TI}: Open traffic intelligence with augmented language model,'' {\em arXiv preprint arXiv:2401.00211}, 2023.

\bibitem{fu2024drive}
D.~Fu, X.~Li, L.~Wen, M.~Dou, P.~Cai, B.~Shi, and Y.~Qiao, ``Drive like a human: Rethinking autonomous driving with large language models,'' in {\em Proceedings of the IEEE/CVF Winter Conference on Applications of Computer Vision}, pp.~910--919, 2024.

\bibitem{lai2023large}
S.~Lai, Z.~Xu, W.~Zhang, H.~Liu, and H.~Xiong, ``Large language models as traffic signal control agents: Capacity and opportunity,'' {\em arXiv preprint arXiv:2312.16044}, 2023.

\bibitem{tang2024large}
Y.~Tang, X.~Dai, and Y.~Lv, ``Large language model-assisted arterial traffic signal control,'' {\em IEEE Journal of Radio Frequency Identification}, 2024.

\bibitem{lopez2018microscopic}
P.~A. Lopez, M.~Behrisch, L.~Bieker-Walz, J.~Erdmann, Y.-P. Fl{\"o}tter{\"o}d, R.~Hilbrich, L.~L{\"u}cken, J.~Rummel, P.~Wagner, and E.~Wie{\ss}ner, ``Microscopic traffic simulation using sumo,'' in {\em 21st international conference on intelligent transportation systems (ITSC)}, pp.~2575--2582, IEEE, 2018.

\bibitem{chu2023survey}
Z.~Chu, J.~Chen, Q.~Chen, W.~Yu, T.~He, H.~Wang, W.~Peng, M.~Liu, B.~Qin, and T.~Liu, ``A survey of chain of thought reasoning: Advances, frontiers and future,'' {\em arXiv preprint arXiv:2309.15402}, 2023.

\end{thebibliography}

\end{document}